# New approaches for CMOS-based devices for large-scale neural recording

Patrick Ruther*, Oliver Paul

Department of Microsystems Engineering (IMTEK), University of Freiburg, Georges-Koehler-Allee 103, 79110 Freiburg, Germany; {paul, ruther}@imtek.de

*Corresponding author

Department of Microsystems Engineering (IMTEK), University of Freiburg, Georges-Koehler-Allee 103, 79110 Freiburg, Germany; ruther@imtek.de; phone +49-761-203-7197

Extracellular, large scale in vivo recording of neural activity is mandatory for elucidating the interaction of neurons within large neural networks at the level of their single unit activity. Technological achievements in MEMS-based multichannel electrode arrays offer electrophysiological recording capabilities that go far beyond those of classical wire electrodes. Despite their impressive channel counts, recording systems with modest interconnection overhead have been demonstrated thanks to the hybrid integration of CMOS circuitry for signal preprocessing and data handling. The number of addressable channels is increased even further by a switch matrix for electrode selection co-integrated along the slender probe shafts. When realized by IC fabrication technologies, these probes offer highest recording site densities along the entire shaft length.

## 1. Introduction

Electrophysiological in vivo recordings of neural activity have greatly helped elucidate aspects of brain function and understand the underlying complex interaction among neurons and between cortical areas. Action potentials, i.e. the basis for neuronal communication, can be accessed extracellularly in the form of single unit activity (SUA), under the condition that microelectrodes implanted into the brain region of interest are positioned close to firing neurons [1]. It has been shown that a large number of individual neurons needs to be simultaneously monitored in order to efficiently tap the communication between brain areas and thus understand brain functionality [1]. Further, in order to extract the neural communication within the vertically organized layers of the cortex [2], a high spatial resolution of extracellular recordings of typically 50 µm and even below over distances of several millimeters is indicated.

Wire electrodes [3] and tetrode configurations thereof [4] have traditionally been applied for in vivo neural recording. The rapid increase in the number and density of recording sites required for large scale recordings is however inhibited by (i) spatial constraints, (ii) an increased complexity of probe implantation and difficulties in precisely positioning different recording sites relative to each other, and (iii) limitations in interconnecting technologies for large numbers of individual wire electrodes. Silicon-based multichannel electrode arrays realized using microelectromechanical system (MEMS) technologies offer a precise arrangement of multiple recording sites building on conceptually different fabrication approaches. As an example, the Utah electrode array (UEA) is based on an array of 10×10 silicon needles carrying a single electrode on each needle tip [5,6]; it is so far the only MEMS-based system with an approval for human trials [7]. In contrast, probes of the Michigan type illustrated in Fig. 1 comprise slender, needle-like probe shafts attached to a broader probe base [8-15]. They carry sets of planar electrodes distributed regularly along the probe shafts; the electrodes are interconnected to contact pads on the probe base by metal lines integrated along the probe shafts.

Both types of silicon-based probe arrays have been applied to extract in vivo extracellular potentials addressing both local field potentials (LFP) from larger neuron populations and single unit activity (SUA) originating from neurons in close proximity to recording sites. In addition, based on the applied materials such as silicon/polysilicon, silicon oxide, silicon nitride, and parylene C for probe body and insulation layers, as well as electrode materials such as gold, platinum, and iridium oxide, an appropriate biocompatibility is offered by both the Michigan-type probes and the Utah array [16,17].

With regards to the perspective of large scale neural recordings, probes of the Michigan type are advantageous as the number of electrodes per shaft is limited only by electrode size, interconnecting line density, and shaft width compatible with tolerable tissue damage. Since these systems can be assembled into truly three-dimensional (3D) probe arrays using a variety of assembly techniques [18-21], brain volumes of application specific geometry can be pervaded by the probes. In contrast, the out-of-plane fabrication process of UEAs has so far prohibited the integration of multiple electrodes per silicon needle. Furthermore, due to the specific fabrication processes of Michigan-type probes, shaft lengths of several 10 mm have been demonstrated [22] enabling electrophysiological recordings from deeper brain structures even in non-human primates. In contrast, the Utah array is limited to shaft lengths of a few millimeters due to the available thicknesses of silicon substrates.

This paper reviews the state of the art of CMOS-based neural probe arrays for in vivo applications building on the Michigan probe concept. It compares the different system architectures that this concept has catalyzed. It discusses opportunities for CMOS circuit integration on the probe base and along the slender probe shafts, and describes most recent probe concepts for in vivo applications in neuroscientific research.

**2. Probe concepts for high density neural recordings**

Silicon probe arrays of the Michigan type are available at different levels of CMOS integration. Passive probes may either apply the hybrid integration with CMOS chips or have the application specific integrated circuit (ASICs) directly integrated in the probe base, as illustrated in Figs. 1(a) or (b), respectively. The next level of CMOS integration takes advantage of circuitry components in the slender probe shanks and/or the probe base, as schematically shown in Figs. 1 (c) and (d).

*2.1 Passive system with hybrid CMOS integration*

Silicon neural probe arrays of the Michigan type have followed an impressive development path over the last four decades. Starting with the pioneering work of K. D. Wise [8], passive silicon probes, i.e., probes without co-integrated circuitry, have been applied in a large number of in vivo studies with different animal models [23-27]. Typically, probes with 16 to 64 channels are interfaced directly either to printed circuit boards (PCB) into so-called acute assemblies, or to highly flexible ribbon cables suitable for long-term applications [20,25,28,29]. Depending on the applied fabrication process [9-11,14], probes with shaft thicknesses between 15 and 100 µm and shaft lengths of up to 40 mm are now commercially available [22,30]. These passive systems are used as one-dimensional (1D) linear probes [23,26], two-dimensional (2D) probe combs [24,25,27] and 3D probe arrays [18,20]. Depending on the assembly approach, systems with up to 512 electrodes have been reported [27].

In view of in vivo applications with freely behaving animals, the increased number of recording sites necessitates dedicated technologies for miniaturized interconnections ensuring the link between the implanted probes and a common, sufficiently small head stage carried by the animal [15,25,27]. The integration of multiplexer chips to merge individual recording channels has helped to reduce the number of interconnections between head stages and external data acquisition systems. Custom-designed CMOS low-power ASICs serve this so-called hybrid approach. These CMOS chips have been directly interfaced to highly flexible ribbon cables [18,20], probe assemblies on PCBs [15,27] as well as to commercial wireless data transmission systems [25]. The ASICs are optimized for the acquisition of both local field potentials (LFP) and single unit activity.

*2.2 COMS Integration on probe base*

In order to relax the laborious effort of assembling probes and head stages, the direct integration of CMOS circuitry on the probe base was first proposed by K. Najafi et al. [31]. This approach is illustrated in Fig. 1(b) for a single shaft probe. Systems with low-noise pre-amplifiers for boosting the neural signals and analog multiplexers for transmitting the data via a small number of output lines have realized this integration concept [32-35]. The probes rely on an adaption of the passive probe fabrication technology to enabling CMOS integration. As with passive silicon probe arrays, the main drawback of this versatile concept is still that each recording site needs to be connected by an interconnection line on the shaft, which limits the number of electrodes per probe shaft. Although the overall promise of the CMOS integration on the probe base is high, it appears that the development of these sophisticated probes is no longer continued by the University of Michigan.

*2.3 CMOS Integration on probe shaft and base*

In order to overcome the restrictions of the existing silicon probes in terms of number and density of the recording sites per shaft, the concept of electronic depth control (EDC) has at first been introduced by the European Integrated Research Project *NeuroProbes* [36]. In this concept CMOS-based multiplexing units are integrated directly on the probe shaft itself [37,38]. The approach allows a number of recording sites to be selected from the large total amount of electrodes distributed densely along the entire probe shaft. As schematically shown in Fig. 1 (c), this affords a dramatic increase in the total number of electrodes compared to existing devices. An additional benefit is that a high recording site density can be achieved over shaft lengths of several millimeters [37,38]. The integrated circuitry of these active CMOS probes has been designed to address subsets of eight recording sites from up to 256 electrodes along probe shafts with lengths of up to 8 mm [37-39]. Among other applications, these probes have served to analyze thalamocortical interactions using a single probe implantation [40].

As reported in further detail in Ref. [37], electrodes are equidistantly arranged in two columns along the probe shafts (Fig. 2(a)). Depending on the applied CMOS technology, electrode pitches of 40.7 µm and 30 µm have been demonstrated, thus enabling probes with different electrode diameters to be obtained without the necessity to redesign the underlying CMOS electronics [37,38]. The integrated circuitry comprises a switch matrix based on transmission gates (TG)[1] controlled by D-type flip-flops (D-FF)[2]. The basic circuit design comprises elementary cells containing four electrodes and eight switching nodes [37]. Each electrode of such an elementary cell can be connected to two out of eight parallel output lines. The modular approach enables multiple elementary cells to be serially arranged along the probe shaft and application specific probe lengths to be easily realized by varying the number of elementary cells per shaft [37]. With the D-FF on a probe shaft connected in a single daisy chain, the resulting shift register allows the TG to be serially programmed using two control lines only, i.e., data in (DIN) and clock (CLK), in combination with two lines for power supply, i.e., VSS[3] and VDD[4] [37]. The corresponding probe base contains only 13 contact pads to address all electrodes along a single shaft, i.e. eight pads for the parallel output lines and five for probe control (DIN, CLK, VSS, VDD, Ground). In the case of comb-shaped probes with four shafts as shown in Fig. 2(a), the number of pads is increased to 44 requiring an area of $2.44 \times 0.68$ mm$^2$ only [37]. As a consequence of this modest number of contact pads, the probe base can be kept small as well; it is thus possible to integrate comb-like EDC probes [37] into slim 3D probe arrays with up to 4096 electrodes covering a total volume of $1.65 \times 1.65 \times 8$ mm$^3$ [41]. Admittedly, only eight electrodes per shaft can be addressed simultaneously. Nevertheless the array offers the freedom to address any electrode along the probe shaft at an electrode pitch down to 30 µm [38].

The most recent high density neural probe array extends the concept of the EDC probes even further. In addition to a switch matrix for the electrodes along the probe shaft, it equips each electrode of the array with a low-noise, low-power amplifier positioned directly under the electrode [42]. The probe contains 455 electrodes, 52 among which can be addressed simultaneously. As shown in Fig. 2(b), sets of 12 circular electrodes in a dual-row arrangement (electrode pitch 35 µm) and a larger square electrode are aligned along the probe shaft. While the circular electrodes serve for recording, local referencing or electrical tissue stimulation may be done by the square electrode of comparatively low electrical impedance. In this probe design, the output of each electrode amplifier can be connected to four different output lines, which allows any combination of four sets of 13 electrodes to be addressed. In contrast to the EDC probes, additional CMOS circuitry is integrated on the probe base. This circuitry carries out signal conditioning tasks. It comprises independently programmable amplification and filtering stages to address SUA and LFP signals for a user-selected group of electrodes. Furthermore, time-division analog multiplexing allows one to route different channels through the same line. A 10-bit successive-approximation-register (SAR) analog-to-digital converter (ADC) further extends the probe functionality [42]. As a consequence of these additional components, the probe base of this single shaft probe occupies an area of $2.9 \times 3.3$ mm$^2$ [42].

### 3. Comparison of probe concepts

---

[1] **Transmission gate** – A transmission gate is type of a relay, i.e. it passes or blocks voltages in both directions depending on a control signal. It comprises two field effect transistors (FET), namely an n-channel and a p-channel metal-oxide-semiconductor (MOS) FET, with the corresponding source and drain contacts being short circuited. The gate terminals of both MOSFETs are connected via a NOT gate, i.e. an inverter. Transmission gates are used to realize CMOS integrated switches or analog multiplexers.

[2] **D-type flip-flop** – Flip-flops are basic building blocks in digital electronics with two stable states. They are used as one-bit memory components triggered with the rising edge of a clock signal. Among the different types of flip-flops, D-type flip-flops are most widely used. They are also known as data or delay flip-flops.

[3] **VSS** describes the positive supply voltage of an integrated circuitry.

[4] **VDD** describes the ground potential of an integrated circuitry.

Due to their successful deployment in neuroscience since many years, passive probe arrays of both the Utah [7,43,44] and Michigan [1,23-27] types have generated massive in vivo recording data. They are applied in large scale neural recording experiments interfacing the passive probes directly with a head stage [1,23-26], or in hybrid [15,27,43,44] and integrated [9,33,34,45] approaches to interface individual electrodes with CMOS-based ASICs. While the number of recording sites along probe shafts is clearly limited by the maximum tolerable shaft width, the separate development of probe arrays without any integrated circuitry and state-of-the-art ASIC chips is a clear advantage of the hybrid integration approach. Still, the overall system including all components to be packed into the head stage occupies several cm$^3$ [27]. This prohibits complete implantation beneficial for experiments with freely behaving animals.

Electrode arrays of the EDC type enable the highest recording site densities over shaft lengths of several millimeters [37,38,46]. This is achieved by CMOS circuitry integrated in the probe shafts. Since the CMOS-based ASIC is limited to the switch matrix connecting electrodes to respective output channels, the probe base can be kept small and thus enables the integration of probe combs into slim 3D probe arrays [37,41]. To date the number of recording sites simultaneously addressable within EDC probes has been eight at maximum. New systems being developed in the framework of the German cluster of excellence *BrainLinks-BrainTools* [47] are pushing the state of the art towards higher channel counts and reduced shaft cross-sections. As an example, advanced EDC probes developed in *BrainLinks-BrainTools* fabricated using a recent 0.18-µm-CMOS technology from X-FAB Semiconductor Foundries AG (Erfurt, Germany) provide 16 parallel output channels for in vivo recording applications as well as two additional stimulation channels at a drastically reduced shaft width of 98 µm.

Furthermore, the EDC approach has been extended by integrating additional CMOS circuitry in the probe shaft, i.e. active pixel amplifiers, and in the probe base, enhancing the system functionality with respect to channel count and interface simplicity [42]. However, increasing the probe base size is expected to prohibit the assembly of the devices into truly 3D probe arrays. For the time being, an output channel count of 52 at minimal probe dimensions represents the current state of the art of CMOS integrated probe arrays. Further extensions aiming to achieve even higher output channel counts well above 250 are currently being addressed in the European research project *NeuroSeeker* (www.neuroseeker.eu).

Table 1 summarizes critical system specifications of proven representative designs, such as probe geometry, channel count, and co-integrated microelectronic functionality. Evident advantages of each approach are partially offset by specific drawbacks. This makes the simple choice between passive probes with a hybrid CMOS integration and CMOS probe arrays with ASICs integrated in the probe shafts or base difficult.

**Acknowledgements**

The authors gratefully acknowledge the financial support by the European Commission in the 6[th] and 7[th] framework program (projects *NeuroProbes* No. IST-027017 and *NeuroSeeker* No. 600925) and the Cluster of Excellence *BrainLinks-BrainTools* funded by the German Research Foundation (DFG, grant number EXC 1086).

*Table 1: Comparison of representative silicon-based probes with advanced CMOS probe arrays used for high density, large-scale neural recordings ($N_{e,s}$: electrodes per shaft; $N_{e,c}$: electrodes per electrode comb or array; $N_S$: number of shafts; $P_e$: electrode pitch, $P_s$: shaft pitch, L: shaft length, W: shaft width, T: shaft thickness).*

| | $N_{e,s}$ {$N_{e,c}$} | $N_s$ | $P_e$ (µm) | $P_s$ (µm) | L/W/T (mm)/(µm)/(µm) | ASIC functionality |
|---|---|---|---|---|---|---|
| **Passive Si probes** | | | | | | |
| Bhandari et al. [6] (UEA) | 1 {100} | 100 | 400 | 400 | 1-1.5/90/90 | - |
| Michigan probes [9,30] | <16 {<128} | 1-16 | >20 | >200 | <10/>50/ ≈15 | - |
| Merriam et al. [20] | 8 {32} | 5×4 | 300, 100 | 200, 125 | 5, 3.4/61/ | |
| Norlin at al. [10] | 32 | 1, 4 | 50-200 | 200-400 | 5/75/20 | - |
| *NeuroProbes* [13,14,26,30] | <32 | 1-4 | >50 | >200 | <40/100-150/ 50-100 | - |
| **Passive probes, hybrid CMOS integration** | | | | | | |
| Harrison et al. [43], Gao et al. [44] (UEA) | 1 {100} | 100 | 400 | 400 | 1-1.5/90t/90t | Amplifier, band pass filter, ADC converter, wireless data transmission |
| Perlin et al. [18] | 4 {16} | 4×4 | 100 | 150 | 4/60/15 | Band width tuning, offset compensation, programmable amplifier gain |
| Fan et al. 0[25] | 7, 4 {15,16} | 2, 4 | 400, 600 | 0.8, 1.2, 1.6 | 7.4, 10.6/140/ 100 | Commercial wireless head stage with preamplifier, filter, multiplexer |
| Berényi et al. [27] | 32 {256} | 8 | 50 | 300 | 5.5/96/15 | Combination with commercial CMOS ASIC from INTAN Inc. |
| **Passive probe active base** | | | | | | |
| Olsson [34] | 8 {32} | 1, 4 | 20-40 | 250 | 5/50/15 | Amplifier, multiplexer, electrical stimulation |
| Csicsvari [32] | 8 {64} | 8 | 20, 50 | 200 | 85/50/12 | Preamplifier |
| **Active shaft, passive base (EDC)** | | | | | | |
| Seidl et al. [37,39,46] | 188 {752} | 1, 4 | 40.7 | 550 | 4/180/50-100 | Switch matrix on shaft; 8/32 electrodes per shaft/comb simultaneously addressable |
| Torfs et al. [38] | 257 {1028} | 1, 4 | 60 | 550 | 8/160750-100 | Switch matrix on shaft; hybrid ASIC with preamplifier and multiplexer; 8/32 electrodes per shaft/comb simultaneously addressable |
| **Active shaft and base** | | | | | | |
| Lopez et al. [42] | 52 {455} | 1 | 35 | - | 10/100/50 | Switch matrix, pixel amplifier, programmable amplifier and filtering stages, multiplexer, ADC |

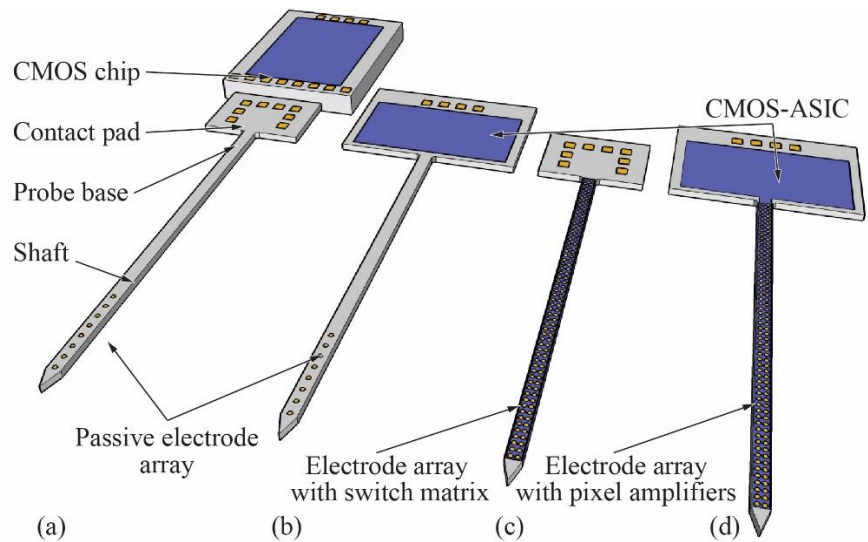

*Figure 1: Schematic of high density probe arrays based on (a) passive silicon probe with CMOS chip in hybrid assembly (wire bonds or cable interfacing probe and chip are not shown), (b) silicon probe with CMOS circuitry on the probe base, (c) EDC probes with a CMOS switch matrix integrated on the probe shaft, and (d) fully CMOS enhanced probe array with integrated circuitry on probe shaft and base.*

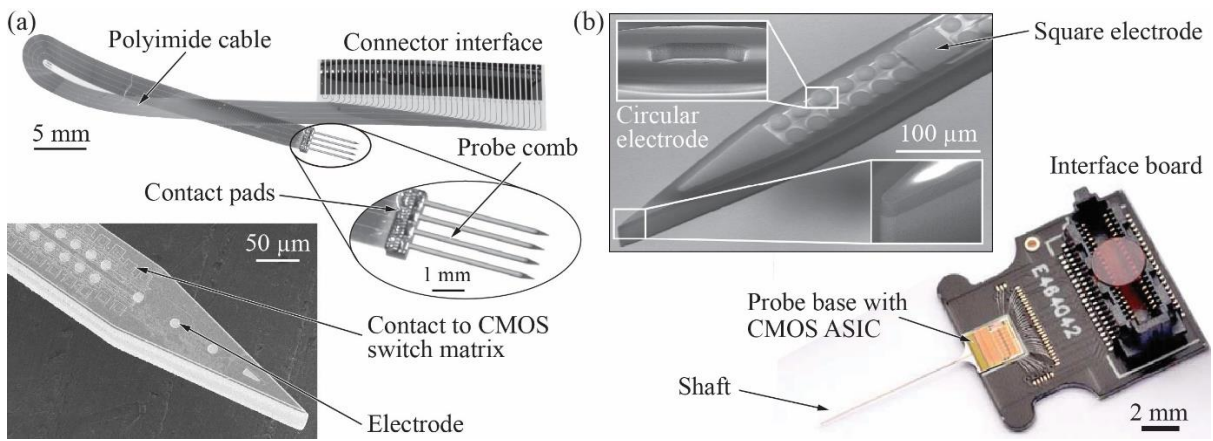

*Figure 2: Scanning electron and optical micrographs of (a) EDC probes with shafts integrated switch matrix as single shaft and comb-shaped array (picture adapted from [35]) and (b) fully CMOS-integrated silicon probe with CMOS circuitry integrated on both probe shaft and base (picture adapted from [38]).*

**References**

[1]  Buzsáki G: **Large-scale recording of neuronal ensembles**. *Nat Neurosci* 2004, **7**(5): 446–451

[2]  V. B. Mountcastle: **The columnar organization of the neocortex**. *Brain* 1997, **120**(4), 701–722

[3]  Nicolelis MAL (Ed): *Methods for Neural Ensemble Recordings* (Frontiers in Neuroscience), 2nd ed. Boca Raton, FL: CRC Press, Dec. 2007

[4]  Wilson MA, McNaughton BL: **Dynamics of the hippocampal ensemble code for space**. *Science* 1993, **261**(5124): 1055–1058

[5]  Campbell P, Jones K, Huber R, Horch K, Normann R: **A siliconbased, three-dimensional neural interface: Manufacturing processes for an intracortical electrode array**. *IEEE Trans Biomed Eng* 1991, **38** (8), 758–768

[6]  Bhandari R, Negi S, Rieth L, Normann RA, Solzbacher F: **A novel masking method for high aspect ratio penetrating microelectrode arrays.** J Micromech Microeng 2009, **19**: 035004

[7]  Hochberg L, Serruya M, Friehs G, Mukand J, Saleh M, Caplan A, Branner A, Chen D, Penn R, Donoghue J: **Neuronal Ensemble Control of Prosthetic Devices by a Human with Tetraplegia.** *Nature* 2006, **442**: 164–171

[8]  Wise KD, Angell JB, Starr A: **An integrated-circuit approach to extracellular microelectrodes.** *IEEE Trans Biomed Eng* 1970, **17**: 238–247

*[9]  Wise K, Sodagar A, Yao Y, Gulari M, Perlin G, Najafi K: **Microelectrodes, Microelectronics, and Implantable Neural Microsystems**. *Proc IEEE* 2008, **96**: 1184–1202

*This paper describes the probe concept of the University of Michigan with respect to silicon-based neural probes comprising a CMOS ASIC integrated in the probe base. It further details the hybrid integration of CMOS chips with passive probes using highly flexible polyimide cables. Similar concepts are described with a 3D probe array in Ref. [18].*

[10] Norlin P, Kindlundh M, Mouroux A, Yoshida K, Hofmann U: **A 32-Site Neural Recording Probe Fabricated by DRIE of SOI Substrates.** *J Micromech Microeng* 2002, **12**: 414–419

[11] Cheung KC, Djupsund K, Dan Y, Lee LP: **Implantable multichannel electrode array based on SOI technology.** J Microelectromech Syst 2003, **12**: 179–184

[12] Herwik S, Kisban S, Aarts A A A, Seidl K, Girardeau G, Benchenane K, Zugaro MB, Wiener SI, Paul O, Neves HP, Ruther P: **Fabrication technology for silicon-based microprobe arrays used in acute and sub-chronic neural recording.** J Micromech Microeng 2009, **19**: 074008 (11pp)

[13] Ruther P, Herwik S, Kisban S, Seidl K, Paul O: **Recent progress in neural probes using silicon MEMS technology**. *IEEJ Trans* 2010, **5**(5): 505–515

[14] Herwik S, Paul O, Ruther P: **Ultrathin Silicon Chips of Arbitrary Shape by Etching Before Grinding.** *J Microelectromech Syst* 2011, **20**: 791–79

*[15]   Du J, Blanche TJ, Harrison RR, Lester HA, Masmanidis SC: **Multiplexed, High Density Electrophysiology with Nanofabricated Neural Probes.** *PLoS ONE* 2011, **6**: e26204

*The paper describes state-of-the-art passive probes with the highest interconnecting line density realized so far. The probe is interfaced with a custom-made ASIC in a hybrid assembly. The challenge of probe interconnection using wire-bonding is reflected in the rectangular ASIC geometry of $8.2 \times 3.5$ mm$^2$.*

[16] Grand L, Wittner L, Herwik S, Göthelid E, Ruther P, Oscarsson S, Neves H, Dombovári B, Csercsa R, Karmos G, Ulbert I: **Short and long term biocompatibility of NeuroProbes silicon probes.** *J. Neuroscience Methods* 2010, **189**: 216-229

[17] Schmidt S, Horch K, Normann R: **Biocompatibility of silicon-based electrode arrays implanted in feline cortical tissue.** *J Biomed Mater Res* 1993, 27(11): 1393-1399

[18] Perlin G, Wise K: **An Ultra Compact Integrated Front End for Wireless Neural Recording Microsystems.** *J Microelectromech Syst* 2010, **19**: 1409–1421

[19] Aarts AAA, Srivannavit O, Wise KD, Yoon E, Puers R, Hoof CV, Neves HP: **Fabrication technique of a compressible biocompatible interconnect using a thin film transfer process.** *JMicromech Microeng* 2011, **21**: 074012

[20] Merriam M, Dehmel S, Srivannavit O, Shore S, Wise K: **A 3-D 160-Site Microelectrode Array for Cochlear Nucleus Mapping.** *IEEE Trans Biomed Eng* 2011, **58**: 397–403

[21] Cheng M-Y, Je M, Tan KL, Tan EL, Lim R, Yao L, Li P, Park W-T, Phua EJR, Gan CL, Yu A: **A low-profile three-dimensional neural probe array using a silicon lead transfer structure.** J Micromech Microeng 2013, **23**: 095013

[22] ATLAS Neuroengineering, Leuven, Belgium

[23] Blanche TJ, Spacek MA, Hetke JF, Swindale NV: **Polytrodes: High-Density Silicon Electrode Arrays for Large-Scale Multiunit Recording.** *J Neurophysiology* 2005, **93**: 2987–3000

[24] Sirota A, Montgomery S, Fujisawa S, Isomura Y, Zugaro M, and Buzsaki G: **Entrainment of neocortical neurons and gamma oscillations by the hippocampal theta rhythm.** *Neuron* 2008, **60**: 683–697

[25] Fan D, Rich D, Holtzman T, Ruther P, Dalley JW, Lopez A, Rossi MA, Barter JW, Salas-Meza D, Herwik S, Holzhammer T, Morizio J, Yin HH: **A Wireless Multi-Channel Recording System for Freely Behaving Mice and Rats.** *PLoS ONE* 2011, **6**: e22033

[26] Bonini L, Maranesi M, Livi A, Fogassi L, Rizzolatti G: **Space-Dependent Representation of Objects and Other's Action in Monkey Ventral Premotor Grasping Neurons.** *J Neuroscience* 2014, **34**: 4108–4119

*[27]   Berényi A, Somogyvári Z, Nagy A J, Roux L, Long J D, Fujisawa S, Stark E, Leonardo A, Harris TD, Buzsáki G: **Large-scale, high-density (up to 512 channels) recording of local circuits in behaving animals.** *J Neurophysiol* 2014, **111** (5), 1132-1149, DOI: 10.1152/jn.00785.2013

*Large scale neural recording using state-of-the-art passive probes with a high channel count up to 256 electrodes per comb-shaped probe in a hybrid assembly with commercially available ASICs are described in this paper.*

[28] Hetke J, Lund J, Najafi K, Wise K, Anderson D: **Silicon Ribbon Cables for Chronically Implantable Microelectrode Arrays.** *IEEE Trans Biomed Eng* 1994, **41**: 314–321

[29] Kisban S: **Silicon-Based Neural Devices: Packaging and Application.** University of Freiburg, Department of Microsystems Engineering, 2010

[30] NeuroNexus Technologies, Ann Arbor, MI, USA

[31] Najafi K, Wise K: **An Implantable Multielectrode Array with On-Chip Signal Processing.** *IEEE J Solid-State Circuits* 1986, **SC-21**: 1035–1044

[32] Csicsvari J, Henze DA, Jamieson B, Harris KD, Sirota A, Barthó P, Wise KD, Buzsáki G: **Massively Parallel Recording of Unit and Local Field Potentials With Silicon-Based Electrodes.** *J Neurophysiology* 2003, **90**: 1314–1323

[33] Olsson R, Wise K: **A three-dimensional neural recording microsystem with implantable data compression circuitry.** *IEEE J Solid-State Circuits* 2005, **40**: 2796–2804

[34] Olsson RH, Buhl D, Sirota A, Buzsaki G, Wise K: **Band-tunable and multiplexed integrated circuits for simultaneous recording and stimulation with microelectrode arrays.** *IEEE Trans Biomed Eng* 2005, **52**: 1303–1311

[35] Sodagar A, Wise K, Najafi K: *A Fully Integrated Mixed-Signal Neural Processor for Implantable Multichannel Cortical Recording.* *IEEE Trans Biomed Eng* 2007, **54**: 1075-1088

[36] Neves HP, Torfs T, Yazicioglu RF, Aslam J, Aarts AA, Merken R, Ruther P, Van Hoof C: **The NeuroProbes project: A concept for electronic depth control.** *Proc IEEE Eng in Med Biol Soc* 2008, 1857

**[37] Seidl K, Herwik S, Torfs T, Neves H, Paul O, Ruther P: **CMOS-Based High-Density Silicon Microprobe Arrays for Electronic Depth Control in Intracortical Neural Recording.** *J Microelectromech Syst* 2011, **20**: 1439–1448

*The paper provides a detailed description of the EDC probe concept realized using a commercially available CMOS process with post-CMOS micromachining based in deep reactive ion etching and grinding (so also [14]). The respective probes are characterized in view of technical performance in Ref. [37] and applied in in vivo recordings in rats described [37, 43]. Similar probes realized using a different CMOS process, are combined with an amplifier and multiplexer ASIC in a hybrid assembly and described in Ref. [38].*

[38] Torfs T, Aarts A, Erismis M, Aslam J, Yazicioglu R, Seidl K, Herwik S, Ulbert I, Dombovari B, Fiath R, Kerekes B, Puers R, Paul O, Ruther P, Van Hoof C, Neves H: **Two-Dimensional Multi-Channel Neural Probes With Electronic Depth Control.** *IEEE Trans Biomed Circ Syst* 2011, **5**: 403–412

[39] Seidl K, Schwaerzle M, Ulbert I, Neves HP, Paul O, Ruther P: **CMOS-Based High-Density Silicon Microprobe Arrays for Electronic Depth Control in Intracortical Neural Recording - Characterization and Application.** *J Microelectromech Syst* 2012, **21**: 1426–1435

[40] Horváth D, Fiáth R, Kereke B P, Dombovári B, Acsády L, Seidl K, Herwik S, Paul O, Ruther P, Neves H P, Ulbert I: **High channel count electrode system to investigate thalamocortical interactions.** *Procedia Computer Science* 2011, **7**: 178–179

[41] Torfs T, Aarts A, Erismis M, Aslam J, Yazicioglu R, Puers R, Van Hoof C, Neves H, Ulbert I, Dombovari B, Fiath R, Kerekes B, Seidl K, Herwik S, Ruther P: **Two-dimensional multi-channel neural probes with electronic depth control.** *IEEE Biomed Circ Syst Conf (BioCAS)* 2010, 198 - 201

**[42] Lopez C, Andrei A, Mitra S, Welkenhuysen M, Eberle W, Bartic C, Puers R, Yazicioglu R, Gielen G: **An Implantable 455-Active-Electrode 52-Channel CMOS Neural Probe.** *IEEE J Solid-State Circuits* 2014, **49**: 248-261

*The paper introduced the most recent linear CMOS-based electrode array which contains – similar to the EDC probes – CMOS circuitry in the probe shank as well as ASICs in the probe base. The probe performance is further improved by CMOS amplifiers positioned under each electrode. The paper provides the probe design, its validation and in vivo application in freely behaving rats*

[43] Harrison RR, Watkins PT, Kier RJ, Lovejoy RO, Black DJ, Greger B, Solzbacher F: **A low-power integrated circuit for a wireless 100-electrode neural recording system.** *IEEE J Solid-State Circuits* 2007, **42** (1): 123–133

[44] Gao H, Walker R, Nuyujukian P, Makinwa KAA, Shenoy K, Murmann B, Meng T: **HermesE: A 96-Channel Full Data Rate Direct Neural Interface in 0.13 μm CMOS.** *IEEE J Solid-State Circuits* 2012, **47**: 1043-1055

[45] Bai Q, Wise K: **Single-Unit Neural Recording with Active Microelectrode Arrays.** *IEEE Trans Biomed Eng* 2001, **48**: 911-920

[46] Dombovári B, Fiáth R, Kerekes B, Tóth E, Wittner L, Horváth D, Seidl K, Herwik S, Torfs T, Paul O, Ruther P, Neves H, Ulbert I: **In vivo validation of the electronic depth control probes.** Biomed Tech 2014, **59(4)**: 283-290

[47] Paul O, Ruther P: **MEMS and more for the brain - THE cluster of excellence BrainLinks-BrainTools at the University of Freiburg.** *Proc IEEE Solid-State Sensor, Actuator and Microsystems Workshop (Hilton Head) 2014*: 1-4